\documentstyle[prd,aps,epsf,eqsecnum,twocolumn,subfigure]{revtex}

\catcode`\@=11

\def\maketitle2{\par 
\begingroup
\let\cite\@bylinecite
\def\thefootnote{\fnsymbol{footnote}}%
\twocolumn[\@maketitle2\vskip2pc]%
\thispagestyle{plain}\@thanks
\endgroup
\def\thefootnote{\arabic{footnote}}%
\setcounter{footnote}{0}%
\let\maketitle2\relax \let\@maketitle2\relax
\let\@thanks\relax \let\@authoraddress\relax \let\@title\relax
\let\@date\relax \let\thanks\relax \let\@abstract\relax 
\let\@pacs\relax}

\def\abstract#1{\gdef\@abstract{{\par 
\bgroup
\ifdim\prevdepth=-1000pt \prevdepth0pt\fi
\hsize\columnwidth
\dimen0=-\prevdepth \advance\dimen0 by17.5pt \nointerlineskip
\small\vrule width 0pt height\dimen0 \relax}{~~}#1\egroup}}

\def\pacs#1{\gdef\@pacs{{\par 
\bgroup
\hsize\columnwidth \parindent0pt
\ifdim\prevdepth=-1000pt \prevdepth0pt\fi
\dimen0=-\prevdepth \advance\dimen0 by20pt\nointerlineskip
\egroup} PACS numbers:~#1}}

\def\preprint#1{\gdef\@preprint{LANL number:~#1}}

\def\@maketitle2{
\@title
\ifdim\prevdepth=-1000pt \prevdepth0pt\fi
\@authoraddress
\@date
\begin{list}{}{\leftmargin=0.10753\textwidth \rightmargin=\leftmargin
\itemsep=1pc\partopsep=-1pc}
\item\@abstract
\item\@pacs; \@preprint

\end{list}
}

\catcode`\@=12

\newcommand{\C}{{\cal C}}

\begin{document}

\title{The quantum roll in d-dimensions and the large-d expansion}
\author{Bogdan Mihaila\thanks{electronic mail:bogdan.mihaila@unh.edu} and
        John~F.~Dawson\thanks{electronic mail:john.dawson@unh.edu}}
\address{Department of Physics, University of New Hampshire,
         Durham, NH 03824}
\author{
   Fred Cooper\thanks{electronic mail:fcooper@lanl.gov},  
   Mary Brewster\thanks{electronic mail:me\_brew@c3serve.c3.lanl.gov},
   Salman Habib\thanks{electronic mail:habib@lanl.gov}
       }
\address{Theoretical Division, MS B285, Los Alamos National
Laboratory, Los Alamos, NM 87545}
\date{\today}

\abstract{ We investigate the quantum roll for a particle in a
$d$-dimensional ``Mexican hat'' potential in quantum mechanics,
comparing numerical simulations in $d$-dimensions with the results of
a large-$d$ expansion, up to order $1/d$, of the coupled closed time
path (CTP) Green's function equations, as well as to a post-Gaussian
variational approximation in $d$-dimensions.  The quantum roll problem
for a set of $N$ coupled oscillators is equivalent to a
$(d=N)$-dimensional spherically symmetric quantum mechanics problem.
For this problem the large-$N$ expansion is equivalent to an expansion
in $1/d$ where $d$ is the number of dimensions.  We use the
Schwinger-Mahanthappa-Keldysh CTP formalism to determine the causal
update equations to order $1/d$.  We also study the quantum
fluctuations $\langle r^2 \rangle$ as a function of time and find that
the $1/d$ corrections improve the agreement with numerical simulations
at short times (over one or two oscillations) but beyond two
oscillations, the approximation fails to correspond to a positive
probability function.  Using numerical methods, we also study how the
long time behavior of the motion changes from its asymptotic ($d
\rightarrow \infty$) harmonic behavior as we reduce $d$. }

\preprint{LAUR 3 98-3411}

\maketitle2 

\section{Introduction}
\label{sec:intro}

The quantum roll problem is the problem of studying the quantum
behavior of a particle starting in an unstable equilibrium at the top
of a potential hill and ``rolling'' down with the constraint $\langle x
\rangle = 0.$ The roll problem recently became relevant to cosmology
with the advent of the new inflation model\cite{ref:qroll}.  In that
model of the early universe one is interested in determining the slow
rollover of the scalar inflaton field in a time evolving semiclassical
gravitational field.  The quantum mechanics of the slow rollover has
been previously studied mostly in perturbation theory or in a mean
field approximation (see however\cite{ref:copeland}).  For the one
dimensional roll problem, it was found that the mean field
approximation broke down before the particle reached the bottom of the
well. Thus it is important to find approximations which are valid for
longer periods of time. One such approximation is to include the $1/N$
corrections to the mean field result.  Our interest in the quantum
mechanical model was due to the fact that for this particular problem
one is able to perform numerical simulations at arbitrary N. This is
due to the radial symmetry of the roll problem. If we were
interested instead in studying tunnelling numerically in an $N$
dimensional anharmonic oscillator, one would be restricted to $N \leq
5$, using even the most powerful computers available\cite{ref:1}.

This paper is constructed as follows: In section \ref{sec:quantumroll}
we state the $d$-dimensional quantum roll problem and discuss how to
choose initial conditions which allow a smooth transition to the $N
\rightarrow \infty$ limit. In section \ref{sec:postgaussian} we derive
a Post-Gaussian variational approximation valid at arbitrary $N$.  In
section \ref{sec:largeN} we review the update equations for the
Green's functions valid up to order $1/N$.  In section
\ref{sec:compisons} we compare both approximations to exact numerical
simulations.  In section \ref{sec:longtime} we study the long time
behavior of the two-point function as a function of $N$. We reserve
the appendices for discussing our numerical methods.

\section{Quantum roll}
\label{sec:quantumroll}

The quantum roll problem for an upside down Harmonic oscillator with
Hamiltonian $H \ = \ ( p^2 - x^2 )/2 $ is the starting point for our
discussion. The solution for the spreading of the wave packet with the
constraint $\langle x \rangle = 0$ is exactly known\cite{ref:qroll}.
For example, if at $t=0$ the wave function is Gaussian, $\psi(x) =
\exp \{ - x^2 / 2 \}$, then for $t > 0$,
\[ 
   \langle x^2(t) \rangle = 1/2 + \sinh^2(t) \>.   
\]
This exponential growth of course does not continue in a bounded
potential.  In this work we will study the quantum roll for a set of
$N$ coupled oscillators in a ``Mexican hat'' potential where the
motion is bounded, as a function of the number of oscillators.  We
take the Hamiltonian to be of the form,
\begin{equation}
   H = - \frac{1}{2} \nabla^2 + V(r)  \>,
\end{equation}
where 
\begin{eqnarray}
   \nabla^2 &=&  \sum_{i=1}^{N} \frac{\partial^2}{\partial x_i^2}
   \>, \qquad 
   r^2 = \sum_{i=1}^{N} x_i^2  \>,
   \nonumber \\ &&
   V(r)     =  \frac{g}{8d} \left ( r^2 \, - \, r_0^2 \right )^2
   \label{eq:potential_r}
   \>. 
\end{eqnarray}
with the Schr\"odinger equation given by:
\begin{equation}
   H \, \Psi(x_i,t) = i \, \frac{\partial \Psi(x_i,t)}{\partial t}
   \>.
\end{equation}
It is convenient to study this problem in a multidimensional
coordinate system with $d \equiv N$, using the radial coordinate $r$
and a set of $d-1$ angular coordinates.  We write the Laplacian
as\cite{ref:Louck,ref:Blaizot}:
\begin{equation}
   \nabla^2 
   \ = \ 
   {\partial^2 \over \partial r^2} 
   \ + \ 
   {(d-1) \over r} {\partial \over \partial r} 
   \ - \ 
   {L^2_{d-1} \over r^2}
   \>,
\end{equation}
where $L^2_{d-1}$ is the generalized orbital angular momentum operator
defined by Louck~\cite{ref:Louck}.  We separate the wave function as
as radial function times a hyperspherical harmonic:
\begin{equation}
   \Psi(x_i,t) \ = \ \Psi_{d, [\lambda] }(r,t) \, 
   Y_{d, [\lambda] }(\Omega)
   \>,
\end{equation}
then the equation for $\Psi_{d, [\lambda] }(r,t)$ becomes:
\begin{equation}
   H(r,l) \Psi_{d, [\lambda] }(r,t) 
   \ = \ i \ {\partial \Psi_{d, [\lambda] }(r,t) \over \partial t}
   \>,
\end{equation}
where
\[
   H(r,l) =
   - \frac{1}{2}
   \left (
      \frac{\partial^2}{\partial r^2}
      + \frac{d-1}{r} \frac{\partial}{\partial r}
   \right ) + {l (l+d-2)  \over r^2} 
   +
   V(r)
   \>.
\]
Here $\lambda$ denotes the set of $(d-1) $quantum numbers, including
the orbital quantum number $l$ needed to specify completely the
hyperspherical harmonic (see for example~\cite{ref:Blaizot}).

For the quantum roll problem the potential is given by
(\ref{eq:potential_r}).  If we start at the top of the hill at $r=0$
with a radially symmetric initial state centered at $r=0$ then there
is no angular momentum and we only have to consider the case $l=0$.
The first order derivative term in the Hamiltonian can be eliminated
by the substitution,
\begin{equation}
   \Phi(r,t) \ = \ r^{(d-1)/2} \, \Psi(r,t)
   \>,
\end{equation}
in which case, the time dependent Schr\"odinger equation for $\Phi$
becomes:
\begin{equation}
    H'(r,l) \  \Phi(r,t) \ = \ i \ {\partial \Phi(r,t) \over \partial t} 
    \>,
\label{eq:redham}
\end{equation}
where
\begin{equation}
   H' \ = \
   - \frac{1}{2} 
      \frac{\partial^2}{\partial r^2} 
   + U(r)  \>,
\end{equation}
with an effective one dimensional potential $U(r)$ given by
\begin{equation}
   U(r) \ = \
   \frac{(d-1)(d-3)}{8 \, r^2} 
   \ + \
   \frac{g}{8d} \left ( r^2 \, - \, r_0^2 \right )^2
   \>. 
\label{eq:Uofr}
\end{equation}
Using this Hamiltonian we can update $\Phi$ using simplectic methods,
or by solving numerically for the eigenfunctions and eigenvalues and
using the expansion:
\begin{equation}
   \Psi(r,t) \ = \ r^{-(d-1)/2} \ \sum_n  \, C_n \, e^{-iE_n t} \, \Phi_n(r)
   \>,
\end{equation}
with $C_n$ being determined from the initial conditions on $\Psi$
using the orthonormality of the $\Phi_n(r)$.

\subsection{Initial conditions}

For most of this paper, we will use for simplicity, Gaussian initial
conditions, since they allow for a simple determination of the $C_n$.
However, in this paper we also want to consider a generalization of
the Gaussian variational method discussed in\cite{ref:pg}, which was
found quite robust in describing the time evolution of pulses in
classical dynamical systems. This generalized wave function has the
ability to approximate in shape an arbitrary spherically symmetric
pulse that is monotonically decreasing around $r=0$. This is
accomplished by adding one more variational parameter $\alpha$ which
changes the shape of the pulse from flat to peaked, including the
Gaussian as the special case $\alpha=1$.  Pulses of this form stay
coherent for a long time in many nonlinear equations and have been
used previously in variational calculations of soliton motion to study
soliton blowup.  They have the advantage of allowing analytical
expressions for the expectation values and thus are ideal for the
variational calculation we will consider below.

Thus we will consider at $t=0$,   normalized wave functions of the form
\begin{eqnarray}
   \Psi_{[\alpha]}(r,0) & = &
   \left [
      \frac{2 \alpha}{\Omega_d \, \Gamma[d/(2\alpha)]}
   \right ]^{1/2}
   \left (
      \frac{2\alpha \, G_{[2 \alpha]}}{d}
   \right )^{-d / (4 \alpha)} \, 
   \nonumber \\ && \times \ 
   \exp \left (
           - \frac{d}{4 \alpha G_{[2 \alpha]}} \, r^{2 \alpha} 
        \right )
   \>,
\label{eq:psi}
\end{eqnarray}
where $\Omega_d$ is the angular volume, $\Omega_d = 2 ~ \pi^{d/2} / \,
\Gamma(d/2)$, such that we satisfy the normalization condition
\begin{equation}
   1 \ = \ 
      \Omega_d \ \int_0^\infty \ 
      | \Psi_{[\alpha]}(r,0) |^2 \ r^{d-1} \ dr
   \>.
\label{eq:norm}
\end{equation}

The energy of this state has the form:
\begin{equation}
   E(\alpha)
   \ = \
      a(\alpha) \, G^{-1 / \alpha}_{[2\alpha]}
      \, + \, e(\alpha) \, G^{2 / \alpha}_{[2\alpha]} 
      \, + \, f(\alpha) \, G^{1/\alpha}_{[2\alpha]}
   \>.
\label{eq:E0}
\end{equation}
where
\begin{eqnarray}
   a(\alpha) & = & 
   \frac{\alpha}{4} \, 
   ( d + 2\alpha - 2) \, 
   \left ( \frac{2 \alpha}{d} \right )^{- 1 / \alpha}
   R(2\alpha - 2, \alpha)
   \>,
\label{eq:a_eqn}   
   \\
   e(\alpha) & = &
   \frac{g}{8d} \, 
   \left ( \frac{2 \alpha}{d} \right )^{2/\alpha}
   R(4,\alpha)
   \>,
\label{eq:e_eqn}   
   \\ 
   f(\alpha) & = & \, -
   \frac{g}{4d} \, r_0^2 \, 
   \left ( \frac{2 \alpha}{d} \right )^{1/\alpha}
   R(2,\alpha)
   \>,
\label{eq:f_eqn}   
\end{eqnarray}
and we have defined $R(\beta,\alpha)$ by:
\begin{equation}
   R(\beta,\alpha)
   \ = \ 
   \Gamma \left ( \frac{\beta + d}{2 \alpha} \right ) \, / \ 
   \Gamma \left ( \frac{d}{2 \alpha} \right )
   \>.
   \label{eq:Rdef}
\end{equation}
In particular, for $\alpha = 1$, the above initial conditions become
\[
   \Psi_{[1]}(r,0) \ = \
   \left [
      \frac{2}{\Omega_d \, \Gamma(d/2)}
   \right ]^{1/2}
   \left (
      \frac{2 G_{[2]}}{d}
   \right )^{-d / 4} 
   e^{{\displaystyle - \frac{d}{4 G_{[2]}} \, r^2}} 
   \>.
\]
We notice that $G_2/d \equiv G$ where 
\[ G = \langle x_i^2 \rangle \]
for each $i$.  Evaluating the energy (\ref{eq:E0}) for the Gaussian
initial state ($\alpha = 1$), we obtain:
\begin{equation}
   E = 
      {g \over 8 d} 
      \left  [ 
         d(d+2) \, G^2 - 2 r_0^2 \, d \, G + r_0^4 
      \right ] + 
      {d \over 8 G}.
\end{equation}

\subsection{Asymptotic form of the wavefunction for large $d$}

The eigenfunctions in the large-$d$ limit are Gaussians times
polynomials, where the Gaussians are centered about the minimum of the
effective potential of the one-dimensonal radial problem.  In order to
get a uniform overlap at arbitrary $d$ it is important to choose
initial conditions so that the expansion in terms of eigenfunctions is
similar at all $d$. This requires us to run the couplings as a
function of $d$ so that the overlap is constant up to terms of order
$1/d^2$.  This can be done in several ways that differ by terms of
order $1/d^2$.  The method presented below leads to uniform results
even at $d=1$ as we change the parameters with $d$.

In order to do this we need to examine the asymptotic form of the wave
function in the large $d$ limit, and choose parameters so that in this
limit, we do not introduce undesirable numerical errors in the initial
decomposition of the wavefunction into eigenmodes.  At $t=0$, the
initial wave function $\Psi_0(r)$ is a normalized gaussian centered
about the origin:
\[
   \Psi_0(r) = {\cal N} \, 
      \exp \left  \{ 
         - \frac{ r^2 }{ 4 G } 
           \right \} \>,
\]
with normalization $\cal{N}$, such that Eq.~(\ref{eq:norm}) is
satisfied.  However the {\em rescaled} wave function $\Phi_0(r)$ at
$t=0$ is given by:
\begin{eqnarray*}
   \Phi_0(r) 
   &=& {\cal N} \, r^{(d-1)/2} \, 
      \exp \left  \{ - \frac{ r^2}{ 4 G} 
           \right \} \>, \\
   &=& {\cal N} \, 
      \exp \left  \{ 
         - \frac{ r^2}{ 4 G} + \frac{d-1}{2} \ln r  
           \right \}  \>.
\end{eqnarray*}
Thus the rescaled wave function, for large $d$, can be approximated by
a gaussian centered about $r = \tilde{r}_{\infty}$.  That is, for
large $d$,
\begin{equation}
   \Phi_0(r) \approx {\cal N} \, 
      \exp \left  \{ 
         - \frac{ (r - \tilde{r}_{\infty})^2 }{ 2 G_\infty } + O(1/d)
           \right \}  \>,
\end{equation}
which defines $\tilde{r}_{\infty}$ and $G_{\infty}$.  For {\em all}
$d$, however, we have:
\begin{eqnarray}
   \tilde{r} = \langle r \rangle_{\!\lower3pt\hbox{$\scriptstyle 0$}}
   & = & {\cal N}^2 \, \Omega_d \, \int_{0}^{\infty} 
      r^{d} \, e^{- r^2 / 2 G} \, {\rm d} r  \>, 
   \nonumber \\
   & = & \sqrt{2 G} \left  [ 
      \frac{ \Gamma( (d+1)/2 ) }{ \Gamma( d/2 ) }
             \right ]  \>. 
   \label{eq:tilder} \\
   \langle r^2 \rangle_{\!\lower3pt\hbox{$\scriptstyle 0$}}
   & = & {\cal N}^2 \, \Omega_d \, \int_{0}^{\infty}
      r^{d+1} \, e^{- r^2 / 2 G} \, {\rm d} r  \>,
   \nonumber \\
   & = & \sqrt{d \, G}  
   \>.
\end{eqnarray}
Thus, we define:
\[
   \frac{G_{\infty}}{2} = 
      \langle r^2 \rangle - \langle r \rangle^2 
      = G 
      \left  \{ 
         d - 
         2 \left  [ 
            \frac{ \Gamma( (d+1)/2 ) }{ \Gamma( d/2 ) }
           \right ]^2  
      \right \} \>,
\]
or
\begin{equation}
   G = \frac{G_{\infty}}
            { 2 d - 
         4 \left  [ \displaystyle
            \frac{ \Gamma( (d+1)/2 ) }{ \Gamma( d/2 ) }
           \right ]^2  } \>. 
   \label{eq:GGinfty}
\end{equation}
In the limit when $d$ goes to infinity, we have
\[
   G(d) \ \rightarrow \ G_\infty \>, \qquad
   \tilde r(d) \ \rightarrow \ \tilde r_\infty 
      \ = \ \sqrt{(d-1) G_\infty}
   \>.
\] 
Now the solution $\Phi(r,t)$ satisfies Eq.~(\ref{eq:redham}) with the
potential function $U(r)$ given by Eq.~(\ref{eq:Uofr}).  In the
spectral method, we expand solutions of Schr\"odinger's equation in
eigenvectors of this equation.  For large $d$, these eigenvectors are
centered about the minimum of the potential $U(r)$.  $U(r)$ has the
expansion,
\begin{equation}
   U(r) = U(\bar{r}) + 
      \frac{1}{2} \bar{m}^2 \, ( r - \bar{r} )^2 + \cdots
   \>,
   \label{eq:Uofr_exp}
\end{equation}
where $\bar{r}$ is given by the solution of the equation,
\begin{equation}
   \frac{(d-1)(d-3)}{4 \, \bar{r}^4} 
   = \frac{g}{2 \, d} \, ( \bar{r}^2 - r_0^2 )  \>,
   \label{eq:barrmin}
\end{equation}
and $\bar{m}^2$ by:
\begin{eqnarray}
   \bar{m}^2 
   &=& \frac{3 \, (d-1)(d-3)}{4 \, \bar{r}^4 } + 
      \frac{g}{2 \, d} \, (3 \bar{r}^2 - r_0^2 ) \>, 
   \label{eq:barmr0} \\
   &=& \frac{g}{d} \, ( 3 \bar{r}^2 - 2 r_0^2 )  \>.
   \nonumber 
\end{eqnarray}
We want to make sure that the difference $\delta r$ between $\bar{r}$
and $\tilde{r}$,
\[
   \delta r = \bar{r} - \tilde{r} \>, 
\] 
remain a constant for all $d$.  This will insure that the overlap
integrals for initial values of the coefficients in a secular
expansion will remain approximately the same for all values of $d$,
and thus numerical errors associated with the initial conditions will
not effect our results.

We also define $m^2$ to be the second derivative of $V(r)$ evaluated
at $r = \bar{r}$,  
\begin{equation}
   m^2 = \frac{d^2 V(r)}{d r^2} \bigg |_{r = \bar{r}} 
       = \frac{g}{2 \, d} \, ( 3 \bar{r}^2 - r_0^2 ) \>,
   \label{eq:mtr0}
\end{equation}
then, (\ref{eq:barmr0}) becomes:
\begin{equation}
   m^2(d) = \bar{m}^2 - \frac{ 3 \, (d-1)(d-3) }{ 4 \, \bar{r}^4 }  \>.
   \label{eq:mtwo}
\end{equation}
Solving (\ref{eq:mtr0}) for $r_0$, substituting into
(\ref{eq:barrmin}) and solving for $g$ gives:
\begin{equation}
   g(d) = \frac{d}{ \bar{r}^2 } 
      \left  \{ 
         \bar{m}^2 - \frac{ (d-1)(d-3) }{ \bar{r}^4 } 
      \right \}  \>.
   \label{eq:grunning}
\end{equation}
Holding $\bar{m}$ fixed means that the frequency of the oscillation
remains the same for all $d$.  Holding $\delta r$ fixed, means that
the overlap between the initial wave function and initial starting
values for the solution remains constant for all $d$.  Thus for {\em
fixed} values of $\delta r$, $G_{\infty}$ and $\bar{m}^2$,
Eqs.~(\ref{eq:tilder}), (\ref{eq:GGinfty}), (\ref{eq:mtwo}), and
(\ref{eq:grunning}), determine values for $G$, $\tilde{r}$, $\bar{r}$,
and $g$ for all values of $d$.  Thus in order to keep the same
accuracy in our solutions, we need to run the coupling constant $g(d)$
and the initial width $G(d)$ as defined above.

We define $\rho_0(d)$ and $\tilde{\rho}(d)$ by the ratios, 
\[
   \rho_0(d) = r_0^2 / ( G d ) \>, \qquad  
   \tilde{\rho}(d) = \tilde{r}^2 / (G d) \>, 
\]  
where $r_0^2$ is calculated from:
\begin{equation}
   r_0^2  = \bar{r}^2 \left  \{
      1 + \frac{1}{2} \left  [ 
         1 - \frac{\bar{m}^2 \bar{r}^4}{(d-1)(d-3)}
                      \right ]^{-1} 
                          \right \} \>.
\end{equation}
Note that in the limit when $d$ goes to infinity, the various
parameters discussed in this section have the limit
\begin{eqnarray}
   g(d) \rightarrow \ 
   g(\infty) \ &=& \ \frac{1}{G_\infty} \
                  \left ( \bar m^2 \, - \, \frac{1}{G_\infty^2} \right ) 
   \\
   m^2(d) \ \rightarrow \ 
   m^2(\infty) \ &=& \ \bar m^2 \, - \, \frac{3}{4 G_\infty^2}
   \\
   \rho_0(d) \ \rightarrow \
   \rho_0(\infty) \ &=& \ 
   1 \, + \, \frac{1}{2 \, (1 - \bar m^2 G_\infty^2)} 
   \label{eq:rho0} \\
   \tilde{\rho}(d) \ \rightarrow
   \tilde{\rho}(\infty) \ &=& \ 
   1
\end{eqnarray}

\begin{figure}
   \centering
   \epsfxsize = 3.in
   \subfigure[$G_\infty = 1$, $\delta r = 2$, and $\bar{m}^2 = 2$.]
             {\epsfbox{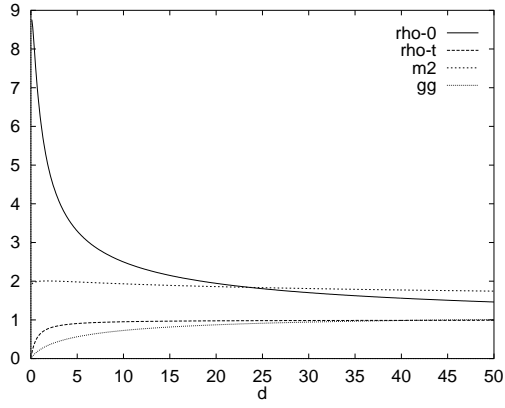}}
   \epsfxsize = 3.in
   \subfigure[$G_\infty = 1$, $\delta r = 2$, and $\bar{m}^2 = 20$.]
             {\epsfbox{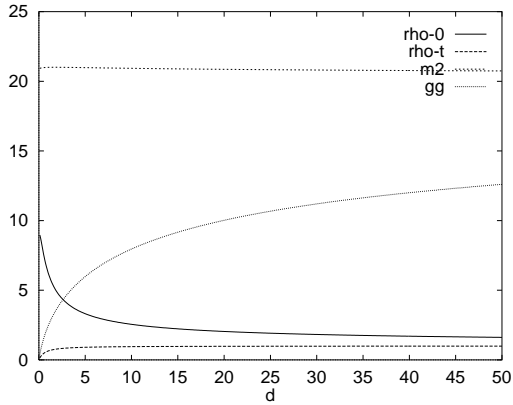}}
   \caption{Plots of $\rho_0$, $\tilde{\rho}$, $g$, and $m^2$
             {\em vs} $d$, for two sets of parameters.}
   \label{fig:roll}
\end{figure}

We show in Fig.~\ref{fig:roll} a plot of $\rho_0(d)$,
$\tilde{\rho}(d)$, $m^2(d)$, and $g(d)$ for the cases $G_\infty = 1$,
$\delta r = 2$, and $m^2 = 2$ and $m^2 = 21$.  The corresponding
initial values are given in Table~\ref{table:table1} and
\ref{table:table2}.  These plots show how these parameters flow as a
function of $d$.  This insures that the initial Gaussian wave function
would have finite overlap with the infinite $d$ eigenfunctions.  Other
choices which differ by terms of order $1/d^2$ are also possible.

If we are using the spectral method for solving the Schrodinger
equation, we need to make sure that the energy of our initial wave
packet is not too large compared with the ground state, so that the
solution will be valid for long times.  In order to estimate the
ground state energy it is sufficient to make a harmonic approximation
around the minimum of the effective one dimensional potential $U(r)$,
given by Eq.~(\ref{eq:Uofr_exp}).  The ground state wave function is
\begin{equation}
   \Phi(r) \ = \ 
   N \ \exp \left [ -\frac{\bar m}{2} \, (r - \bar r)^2 \right ]
   \>,
\end{equation}
where $N$ is a normalization constant.  Correspondingly, the ground
state energy is approximately
\begin{equation}
   E_0 \ = \
   U(\bar r) \ + \ \bar m / 2
   \>.
\end{equation}
We used this result to get a feeling for how to choose our initial
conditions.  Afterward we determined our parameters from the
asymptotic relations, keeping this result in mind.

\section{Time-dependent Post-Gaussian variational method}
\label{sec:postgaussian}

Because of the radial symmetry of the quantum roll problem, it is
possible to make a simple generalization of the Gaussian approximation
which allows us to track wave packets which are have arbitrarily high
correlation functions and thus might be more coherent in their
behavior in a non-linear potential.  These wave packets were used
previously\cite{ref:pg} to study soliton behavior and blowup in the
classical nonlinear Schr\"odinger equation in arbitrary $d$.  These
wave functions have the property that at $d=1$ they give a much better
ground state wave function in that the wave function becomes quite
flat and can span both wells.  At large $d$ this wave function goes
over to the gaussian limit since that is the correct ground state at
large $d$.  Since the analytic large-d expansion is an expansion about
a Gaussian, we expect the post-Gaussian variational wave to be
compatible with that expansion only at large $d$.  At small $d$, if we
use the parameters obtained from the ground state energy, our initial
state will be quite different from a Gaussian and not track an initial
Gaussian very well in detail.  We can, however start with non-Gaussian
initial conditions and make direct comparisons numerically.

The time dependent Schr\"odinger equation can be obtained by varying
the Dirac action
\begin{eqnarray}
   \Gamma & = &
   \int {\rm d}t \,
       \langle \Psi(t) \, | \, 
          i  \partial / \partial t - H \, 
       | \, \Psi(t) \rangle
   \>.
\end{eqnarray}
The Schr\"odinger equation results by asserting that the action
$\Gamma$ be stationary against arbitrary variations of the wave
function $\Psi(r,t)$.  We approximate the {\em true} wave function
$\Psi(r,t)$ using a generalized Gaussian trial function
of~\cite{ref:pg}
\begin{eqnarray}
   \Psi_v(r,t) & = &
   \left [
      \frac{2 \alpha}{\Omega_d \, \Gamma[d/(2\alpha)]}
   \right ]^{1/2}
   \left (
      \frac{2\alpha \, G_{[2 \alpha]}}{d}
   \right )^{-d/(4 \alpha)} \, 
   \nonumber \\ && \times \ 
   \exp \left [
           - r^{2 \alpha} \left ( \frac{d}{4 \alpha G_{[2 \alpha]}}
                                  - i \Lambda
                          \right )
        \right ]
   \>.
\end{eqnarray}
Correspondingly, the expectation value of $r^\beta$ is 
\begin{eqnarray}
   \langle r^\beta \rangle
   & = & 
   \left ( 
      \frac{2 \alpha \, G_{[2 \alpha]}}{d} 
   \right )^{\beta / (2 \alpha)}
   R(\beta,\alpha)
   \>,
\label{eq:rbeta}   
\end{eqnarray}
where $R(\beta,\alpha)$ is given by Eq.~(\ref{eq:Rdef}).
In particular for $\beta = 2\alpha$,
\[
   G_{[2 \alpha]} = \langle r^{2 \alpha} \rangle
   \>.
\]
Note that in ref.~\cite{ref:pg}, we considered only the case when
$\alpha=1$ where 
\[ 
   G_{[2]} \ = \ \langle r^2 \rangle \ = \ d \, G 
   \>.
\]
Using Eq.~(\ref{eq:rbeta}), we can calculate the expectation value of
the Hamiltonian, $\langle H \rangle = \langle H_{ {\rm free} } \rangle
+ \langle H_{ {\rm int} } \rangle $, with
\begin{eqnarray}
   \langle H_{ {\rm free} } \rangle 
   & = & 
   - \frac{\Omega_d}{2} \times 
   \nonumber \\
   && \>  
   \int  {\rm d}r~ r^{d-1}
      \Psi_v^*(r,t) \left ( \frac{\partial^2}{\partial r^2} 
                       + \frac{d-1}{r} \frac{\partial}{\partial r}
               \right ) 
      \Psi_v(r,t)  
   \nonumber \\
   & = &
   \alpha \, 
   ( d + 2\alpha - 2) \, 
   \left ( \frac{d}{4\alpha G} - i \Lambda \right )
   \langle r^{2 \alpha - 2} \rangle
   \nonumber \\ &&
   - \, 2 \alpha^2 \, 
   \left ( \frac{d}{4\alpha G} - i \Lambda \right )^2
   \langle r^{4 \alpha - 2} \rangle
   \nonumber \\
   & = &
   \alpha \, 
   ( d + 2\alpha - 2) \, 
   \left [ 
      \left ( \frac{d}{4\alpha G} \right )^2 
      + \Lambda^2 
   \right ]
   \nonumber \\ &&
   \left ( \frac{2\alpha G}{d} \right )^{2 - \frac{1}{\alpha}}
   \Gamma_{(2\alpha - 2, \alpha)}
\end{eqnarray}
and
\begin{eqnarray}
   \langle H_{ {\rm int} } \rangle
   & = & 
   \Omega_d  \int 
      \Psi_v^*(r,t) \ V(r) \ \Psi_v(r,t) \, r^{d-1}\, {\rm d}r
   \nonumber \\
   & = &
   \frac{g}{8 d} \, \langle r^4 \rangle 
   - \frac{g}{4 d} \, r_0^2 \, \langle r^2 \rangle
   + \frac{g}{8d} \, r_0^4
   \>.
\end{eqnarray}
We also find the expectation value of $i\, \partial / \partial t$ as
\begin{eqnarray}
   \langle i \, \partial / \partial t \rangle 
   & = & 
   \frac{i}{2} \,\Omega_d  
      \int \
      \left [ \Psi_v^* \frac{\partial \Psi_v}{\partial t}
              - \frac{\partial \Psi_v^*}{\partial t} \Psi_v
      \right ] \, r^{d-1}\, {\rm d}r
   \nonumber \\ & = & 
   - \ G_{[2\alpha]} \, \dot \Lambda
   \>,
\end{eqnarray}
so the effective action can be written as:
\begin{eqnarray}
   && \Gamma = 
   \int {\rm d}t \ 
   \bigl \{
      - \, G_{[2\alpha]} \, \dot \Lambda 
      \, - \, a(\alpha) \, G_{[2\alpha]}^{  - 1 / \alpha}
   \nonumber \\ && \quad
      \, - \, b(\alpha) \, G_{[2\alpha]}^{2 - 1 / \alpha} \Lambda^2 
      - \, e(\alpha) \, G_{[2\alpha]}^{2 / \alpha} 
      \, - \, f(\alpha) \, G_{[2\alpha]}^{1/\alpha}
   \bigr \}
   \>,
\end{eqnarray}
where the coefficients $a(\alpha)$, $e(\alpha)$ and $f(\alpha)$ are
given by Eqs.~(\ref{eq:a_eqn}, \ref{eq:e_eqn}, \ref{eq:f_eqn}).  In
addition we have introduced
\begin{eqnarray}
   b(\alpha) & = & a(\alpha) \,
   \left ( \frac{4\alpha}{d}  \right )^2
   \>.
\label{eq:b_eqn}   
\end{eqnarray}

The variational equations of motion are obtained from
\begin{eqnarray}
   \frac{\delta \Gamma}{\delta \Lambda} 
   & = &
   \dot G_{[2\alpha]} 
   \, - \, 2 b(\alpha) \, G_{[2\alpha]}^{2 - 1 / \alpha} \Lambda
   = 0  \>,
\label{eq:lambda}
   \\ 
   \frac{\delta \Gamma}{\delta G_{[2\alpha]}}
   & = &
   - \dot \Lambda 
   + \frac{a(\alpha)}{\alpha} G_{[2\alpha]}^{-1 - 1/\alpha}
   \nonumber \\ &&
   - b(\alpha) \left ( 2 - \frac{1}{\alpha} \right ) 
     G_{[2\alpha]}^{1 - 1 / \alpha} \Lambda^2
   \nonumber \\ && 
   - \frac{2 e(\alpha)}{\alpha} G_{[2\alpha]}^{-1 + 2/\alpha} 
   - \frac{f(\alpha)}{\alpha} G_{[2\alpha]}^{-1 + 1/\alpha}    
   = 0  \>.
\label{eq:Geq}
\end{eqnarray}
Note that Eq.~(\ref{eq:lambda}) can be solved for $\Lambda$. 
We get
\begin{equation}
   \Lambda \ = \ 
   \frac{1}{2 b(\alpha)} \, \dot G_{[2\alpha]} \, 
   G_{[2\alpha]}^{-2 + 1 / \alpha}
   \>.
\end{equation}
Since the effective action is stationary against variation of the
trial wave function $\Psi_v(t)$, the equations of motion
(\ref{eq:lambda}, \ref{eq:Geq}) guarantee energy conservation:
\begin{eqnarray}
   E/d & = & 
      a(\alpha) \, G_{[2\alpha]}^{-1 / \alpha}
      \, + \, b(\alpha) \, G_{[2\alpha]}^{2 -1 / \alpha} \Lambda^2 
   \nonumber \\ &&
      \, + \, e(\alpha) \, G_{[2\alpha]}^{2 / \alpha} 
      \, + \, f(\alpha) \, G_{[2\alpha]}^{1/\alpha}
   \>.
\label{eq:E_har}   
\end{eqnarray}
The conservation provides a first integral of the motion and gives us
a non-linear differential equation to solve in terms of $G_{[2\alpha]}$
and the initial energy.

Stationary solutions are obtained by requiring $\dot G_{[2\alpha]} =
\dot \Lambda = 0$.  Then, from Eqs.~(\ref{eq:lambda}, \ref{eq:Geq}) we
obtain $\Lambda = 0$ and an equation for $G_{[2\alpha]}$
\begin{equation}
   2 e(\alpha) \, G_{[2\alpha]}^{3/\alpha} 
   \, + \, f(\alpha) \, G_{[2\alpha]}^{2/\alpha}    
   \, - \, a(\alpha) 
   \ = \ 0
   \>.
\end{equation}
The energy of this stationary solution is:
\[
   \frac{E(\alpha)}{d} = 
      a(\alpha) \, G_{[2\alpha]}^{-1 / \alpha}
      \, + \, e(\alpha) \, G_{[2\alpha]}^{2 / \alpha} 
      \, + \, f(\alpha) \, G_{[2\alpha]}^{1/\alpha}
   \>.
\]
By minimizing this energy with respect to $G_{[2\alpha]}$ and $\alpha$
we get the best estimate of the ground state wave function for this
class of trial wave functions.  We must remember that we expect the
actual ground state wave function to have its support at $\bar{r}$ and
not around zero.  This generalized approximation has the feature that
as $d \rightarrow \infty$, $\alpha_c \rightarrow 1$.

\begin{figure}
   \epsfxsize = 3.in
   \centerline{\epsfbox{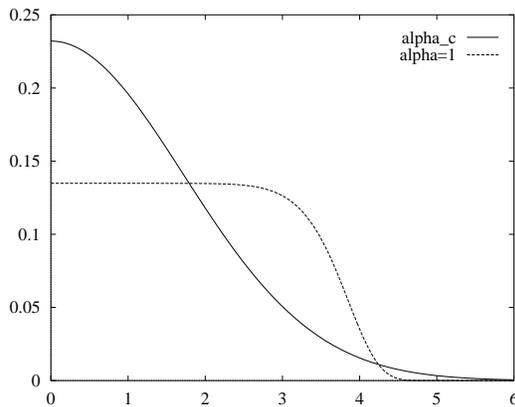}}
   \caption{Gaussian versus generalized Gaussian wave functions, 
            for $g_\infty=1$ and $d=1$ with $r_0 = 2.9359$.}
   \label{fig:psi0psi1}
\end{figure}

First let us look at the ground state wave function at $d=1$,
$g_{\infty} = 1$, which, from Table~\ref{table:table1}, corresponds to
$g = 0.232$ and $r_0 = 2.9359$.  We compare the variational wave
function and ground state energy with the best Gaussian result.
Minimizing the energy with respect to the parameters $G_{[2\alpha]}$
and $\alpha$, we find that the optimal parameters are:
\begin{equation}
   \alpha_c = 5.25774 \>; \quad G_{[2\alpha_c]}= 152039. 
   \>,
\end{equation}
leading to a normalized wave function given by:
\[
   \psi(r) \ = \ 
   0.134818  \ \exp 
      \left  \{ 
         - \, 6.25428 \times 10^{-07} \ r^{10.515483} \, 
      \right \} \>,
\]
corresponding to an energy of $E(\alpha_c) = - 1.05847$.  In
comparison, the Gaussian approximation ($\alpha =1$) leads to the
results:
\begin{equation}
   \alpha= 1.0 \>; \quad G_{[2]} = 2.95542 
   \>, 
\end{equation}
with a wave function given by:
\[
   \psi(r) \ = \
   0.232058  \ \exp 
      \left  \{ 
         - \, 0.16918 \ r^2
      \right \} 
\]
corresponding to an energy of $E(\alpha=1) = -0.67531$.  A comparison
of these two wave functions is found in Fig.~\ref{fig:psi0psi1}.  Thus
the more general wave function has support over both wells and does
considerably better for the energy.

\begin{figure}
   \centering
   \epsfxsize = 2.2in
   \subfigure[$d$=1]
             {\epsfbox{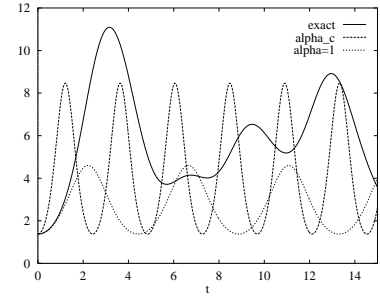}}
   \epsfxsize = 2.2in
   \subfigure[$d$=5]
             {\epsfbox{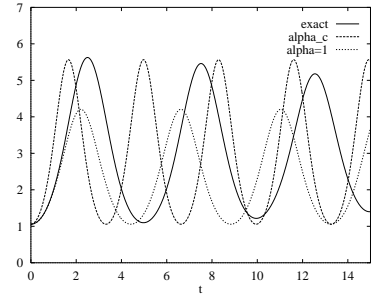}}
   \epsfxsize = 2.2in
   \subfigure[$d$=10]
             {\epsfbox{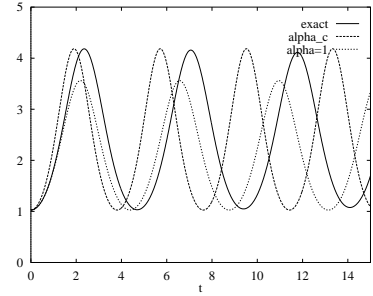}}
   \epsfxsize = 2.2in
   \subfigure[$d$=20]
             {\epsfbox{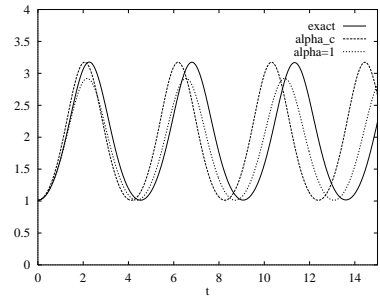}}
   \caption{Gaussian initial conditions with $g_\infty=1$. Here we
   compare the exact results with both the post-Gaussian aproximation
   normalized to $\langle r^2 \rangle$ and the Gaussian approximation.}
   \label{fig:var_alpha}
\end{figure}

\begin{figure}
   \centering
   \epsfxsize = 2.2in
   \subfigure[$d$=1]
             {\epsfbox{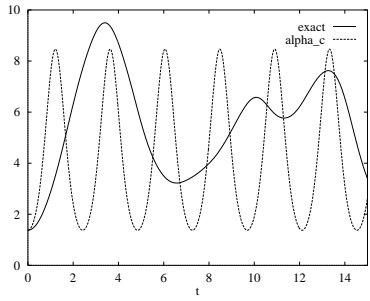}}
   \epsfxsize = 2.2in
   \subfigure[$d$=5]
             {\epsfbox{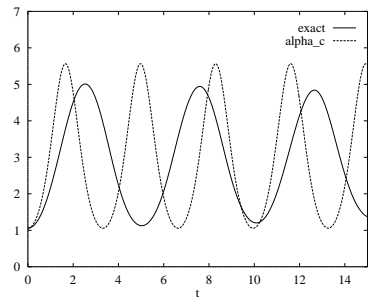}}
   \epsfxsize = 2.2in
   \subfigure[$d$=10]
             {\epsfbox{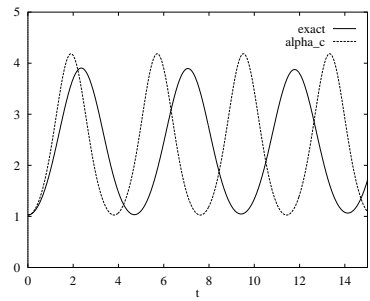}}
   \epsfxsize = 2.2in
   \subfigure[$d$=20]
             {\epsfbox{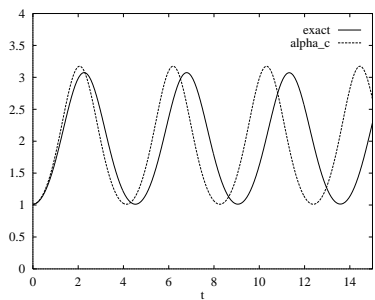}}
   \caption{Post-Gaussian initial conditions with $g_\infty=1$. Here
   we compare the exact results with the Post Gaussian approximation
   with $\alpha_c$ fixed by minimizing the ground state energy.}
   \label{fig:var_alpha_c}
\end{figure}

Because of this wider support, the post-Gaussian approximation does a
better job in reproducing the amplitude of the excursions in the time
dependent problem when compared to the Gaussian approximation even
though it does not represent the early time evolution.  At larger $d >
10$ when $\alpha$ becomes closer to unity then the entire evolution is
modelled better by this type of trial wave function. This is seen in
Fig.~\ref{fig:var_alpha}.  We can also compare the generalized
Gaussian trial wave function result with with a numerical simulation
starting from the non-Gaussian initial conditions using the split
operator technique described in \cite{ref:sp_op}.  The results at two
different $d$'s are shown in Figs.~\ref{fig:var_alpha_c}.  Again we
see that we need $d \ge 10$ for this approximation to give reasonable
results.

\section{large-$N$ expansion}
\label{sec:largeN}

The large-$N$ ($d$) expansion for the $d$-dimensional anharmonic
oscillator is the zero space dimensional limit of the $\phi^4$ field
theory formalism discussed in ref.~\cite{ref:CHKMPA}. The effective
action to order $1/d$ is given by [here we use the field theory
notation of ref.~\cite{ref:CHKMPA}: $d \rightarrow N$, $g \rightarrow
\lambda$, and $\mu^2 \rightarrow - g r_0^2 / (2 d)$],
\begin{eqnarray}
   \lefteqn{\Gamma[\chi,x_i] = } 
   \nonumber \\
   & &
      \int_{\cal{C}} {\rm d}t \, \Bigl\{
               \frac{1}{2} \sum_i 
         \Bigl\{ 
            \dot{x}_i^2(t) - \chi x_i^2(t) \Bigr\} + 
         \frac{i N}{2} \ln [ G_0^{-1}(t,t) ]
          \nonumber \\
   & & \quad {}+ 
      \frac{N \chi^2(t)}{2\lambda} - \frac{N \mu^2 \chi(t)}{\lambda} 
        + \frac{i}{2} \ln [ D^{-1}(t,t) ]  
                             \Bigr\}
   \>, 
\label{eq:effaction}
\end{eqnarray}
where $G^{-1}_{ab}(t,t')$ and $D^{-1}(t,t')$ are the inverse
propagators for $x_i$ and $\chi$, given ( for $\langle x_i \rangle
=0$) by
\begin{eqnarray}
   G^{-1}_{ab}(t,t') & = & 
      \left  \{ 
         \frac{ {\rm d}^2 }{ {\rm d} t^2 } + \chi(t) 
      \right \} \, \delta_{\C} (t,t') \, \delta_{ab} \nonumber \\
   & \equiv & G_0^{-1}(t,t') \, \delta_{ab} \>, \\
   D^{-1}(t,t') & = &  
      - \frac{N}{\lambda} \delta_{\C}(t,t') 
      - \Pi(t,t')
   \>, \\
   \Pi(t,t') & = & 
      - \frac{i}{2} \, \sum_{a,b=1}^{N} G_{ab}(t,t') \, G_{ba}(t',t) 
   \>.
\end{eqnarray}
Here $\delta_{\C} (t,t')$ is the closed time path delta function. 
The auxilliary variable $\chi(t)$ obeys the constraint equation given by
\begin{equation}
   \chi(t) =  \mu^2 +  
   \frac{\lambda}{2 N} \sum_{a=1}^{N} 
      \frac{1}{i} {\cal G}_{aa}(t,t) \>,
\label{eq:Chieqn}
\end{equation}
and where the full $x_i$ propagator ${\cal G}(t,t')$ and self
energy $\Sigma(t,t')$ to order $1/N$ are given by
\begin{eqnarray}
   &&{\cal G}_{ab}(t,t')
    = 
   G_{ab}(t,t') \label{eq:Gfull}  \\
   && 
   {}-  \sum_{c,d = 1}^{N} 
      \int_{\C} {\rm d}t_1 \, \int_{\C} {\rm d}t_2 \,
      G_{ac}(t,t_1) \, \Sigma_{cd}(t_1,t_2) \, G_{db}(t_2,t') 
   \nonumber \\
   &&\Sigma_{cd}(t,t')
    = 
   i \, G_{cd}(t,t') \, D(t,t')   \>.
   \nonumber
\end{eqnarray}
These equations are also derived in (2.18--2.22) of
ref.~\cite{ref:CHKMPA}.  In order to solve for $D(t,t')$, we first
write
\begin{equation}
   \frac{N}{\lambda} \, D(t,t') \ = \
      - \delta_{\C}(t,t') 
      + \frac{N}{\lambda} \, \Delta D(t,t') 
   \>.
\label{eq:Dsubst}
\end{equation}
Then we find that $\Delta D(t,t')$ satisfies the integral equation,
\begin{equation}
   \Delta D(t,t') 
   \ = \ 
   {\lambda \over N} \, \Pi(t,t')
   - \int_{\C} {\rm d}t'' \, 
            \Pi(t,t'') \, \Delta D(t'',t') 
   \>,
\label{eq:Dtileqn}
\end{equation}
in agreement with (2.13--2.16) of Ref.~\cite{ref:CHKMPA}.
We are now in a position to solve these coupled equations for the
motion of ${\cal G}_{ab}(t,t') $ and $\chi(t)$ for given initial 
conditions. We solve Eq.~(\ref{eq:Chieqn}) simultaniously with
(\ref{eq:Gfull}) and (\ref{eq:Dtileqn}), using the Chebyshev
expansion technique of appendices A and B of ref.~\cite{ref:MDC}.

\section{Comparison of approximations with exact numerical 
   simulations.}
\label{sec:compisons}

\begin{figure}
   \centering
   \epsfxsize = 2.2in
   \subfigure[$d$=1]
             {\epsfbox{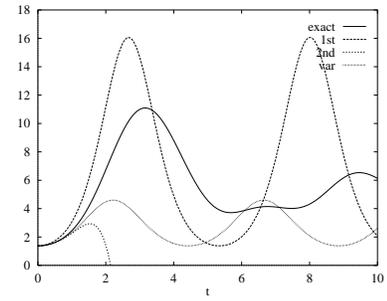}}
   \epsfxsize = 2.2in
   \subfigure[$d$=5]
             {\epsfbox{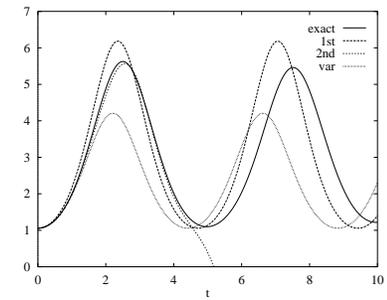}}
   \epsfxsize = 2.2in
   \subfigure[$d$=10]
             {\epsfbox{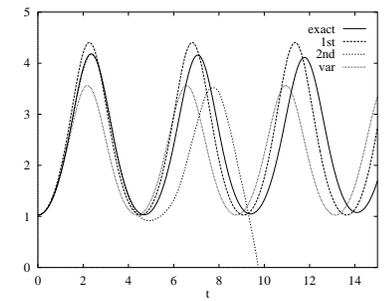}}
   \epsfxsize = 2.2in
   \subfigure[$d$=20]
             {\epsfbox{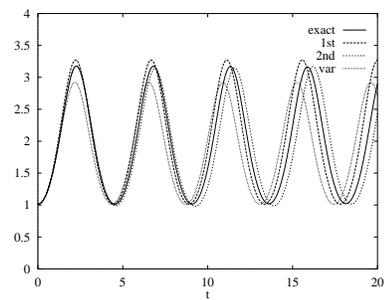}}
   \caption{Gaussian initial conditions with $g_\infty=1$. Here we
   compare exact result with the leading and next to leading
   order large-$N$ approximation and to the Gaussian
   approximation.}
   \label{fig:set1_fig}
\end{figure}

In this section we will compare the exact solution (determined
numerically) with both the variational method and the large-$N$
expansion.  In order to have a smooth $d \rightarrow \infty$
transition, we will allow our parameters to run as a function of $d$
as described above.  We use the set of parameters listed in tables
\ref{table:table1} and \ref{table:table2}, which correspond to the
choice of $G_\infty=1$, $\delta r = 2$, and the two cases $\bar m^2 =
2$, $g_{\infty} =1$ and $\bar m^2 = 21$; $g_{\infty} =20$,
respectively.  First let us discuss Gaussian initial conditions.  In
Figs.~\ref{fig:set1_fig} we present the results for $\angle r^2/d
\rangle$ which correspond to $g_{\infty}=1$.  We find that at short
times (less than 2 oscillations) the calculation which includes $1/N$
corrections tracks the exact result the closest, however when it
starts deviating from the exact result it can lead to a negative
expectation value. This is evidence that the next to leading order
large-N approximation does not correspond to a positive definite
probability function.  At leading order, the large-N approximation is
equivalent to a Gaussian wave function (or density matrix) and all the
expectation values of even moments of operators are positive.  This
problem of not having a positive definite probability associated with
this type of approximation scheme also arises if we consider
approximations which are exact truncations of the Green's function
hierarchy\cite{ref:bett} at the level of the connected four point
function.

\begin{figure}
   \centering
   \epsfxsize = 2.2in
   \subfigure[$d$=1]
             {\epsfbox{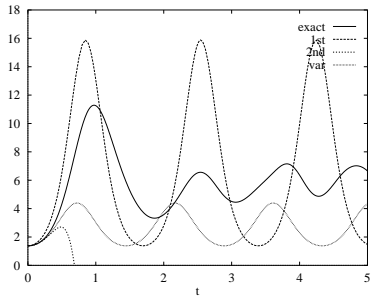}}
   \epsfxsize = 2.2in
   \subfigure[$d$=5]
             {\epsfbox{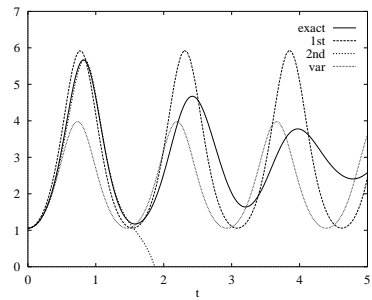}}
   \epsfxsize = 2.2in
   \subfigure[$d$=10]
             {\epsfbox{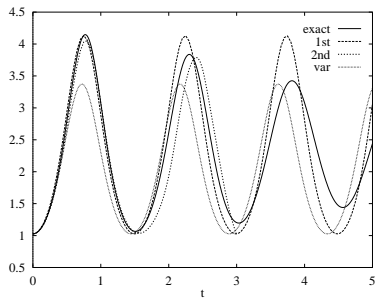}}
   \epsfxsize = 2.2in
   \subfigure[$d$=20]
             {\epsfbox{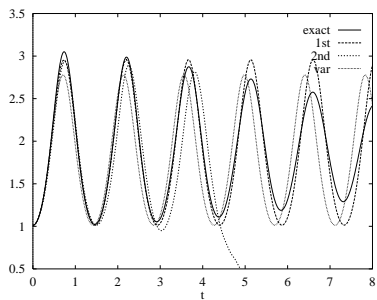}}
   \caption{Gaussian initial conditions with $g_\infty=20$. Similar to
   Fig. 5 we notice that the next to  leading order large-$N$
   approximation becomes unstable for small $d$ at modest time scales.}
   \label{fig:set2_fig}
\end{figure}

The leading order in large-N calculation is closer in amplitude to the
exact result than the Hartree variational approximation.  These latter
two approximations do not suffer from the illness of the next to
leading large-N approximation.  We see that all the approximations
approach the exact one as $d \rightarrow \infty$ as they must.  In
Figs.~\ref{fig:set2_fig} we see the same type of behavior at a larger
value of $g_{\infty}=20$.  The main difference here is that the time
scale for a single oscillation has been reduced, but the behavior in
terms of the oscillation time scale is similar to the previous case.

We have also compared the post-Gaussian variational approximation for
Gaussian initial data.This comparison is shown in
Figs.~\ref{fig:var_alpha}.  Here, the initial states are clearly
different so we have normalized the two wave functions so as to have
the same value for $\langle r^2 \rangle$.  What is interesting here is
that the post-Gaussian approximation gives a better approximation for
the amplitude of the oscillation, which "cures" the previously found
failure of the Gaussian approximation to reach the bottom of the
Mexican Hat potential in one dimension.  To truly test the
post-Gaussian approximation we used non-Gaussian initial conditions
pertaining to the value of $\alpha=\alpha_c$ found by minimizing the
ground state energy.  To obtain our numerical results here, it was not
possible to use the eigenfunction method because of the number of
eigenfunctions required to approximate this type of wave function.
Instead we used the split operator technique of \cite{ref:sp_op}.  As
shown in Figs.~\ref{fig:var_alpha_c} we see that the post-Gaussian
approximation does quite well with the amplitude of the oscillation,
but only at large $d \ge 10$ does it start getting the frequency of
oscillations correct.

\section{Long-Time Behavior of \lowercase{$\langle r^2 \rangle$} 
   as a function of \lowercase{$d$}.}
\label{sec:longtime}

At large $d$ the effective one dimensional radial potential is very
deep and becomes harmonic for oscillations around the minimum.  Thus
as $d\rightarrow\infty$ an initial Gaussian wave packet should stay
gaussian, because the gaussians are coherent states of the harmonic
oscillator.  However, the solutions for finite dimension $d$ can not
be expected to maintain the coherence of the original Gaussian state
over long time intervals.  It is of interest to observe the actual
long-time behavior of the solutions that arise from Gaussian initial
conditions when the potential is not purely quadratic.

The method of eigenfunction expansion can be used to calculate the
solution over large time intervals.  The accuracy of the long-term
behavior is limited primarily by the accuracy of the eigenvalues, so
the tolerance of the eigenvalue calculation was reduced until
numerical convergence of the solution was observed over the time
intervals of interest.

\begin{figure}
   \centering 
   \epsfxsize = 2.2in 
   \subfigure[$d$=1]
             {\epsfbox{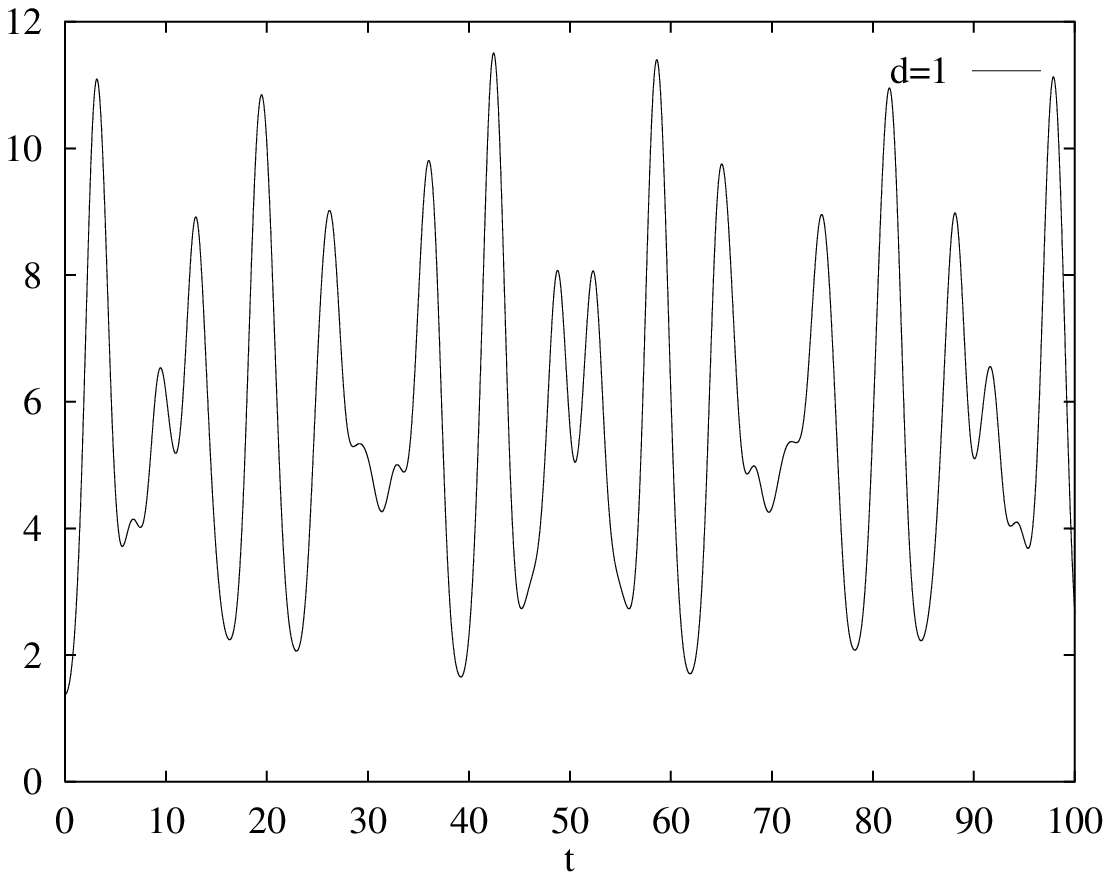}} 
   \epsfxsize = 2.2in 
   \subfigure[$d$=5]
             {\epsfbox{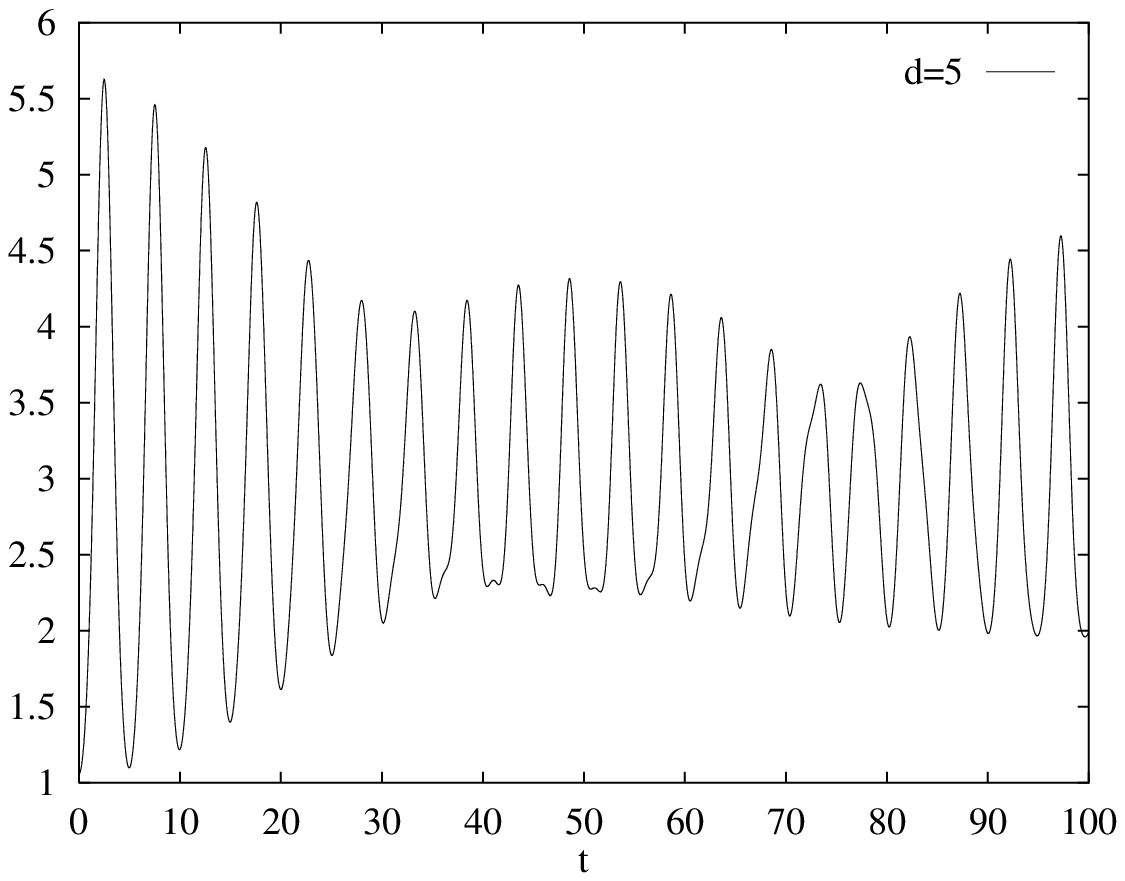}} 
   \epsfxsize = 2.2in 
   \subfigure[$d$=10]
             {\epsfbox{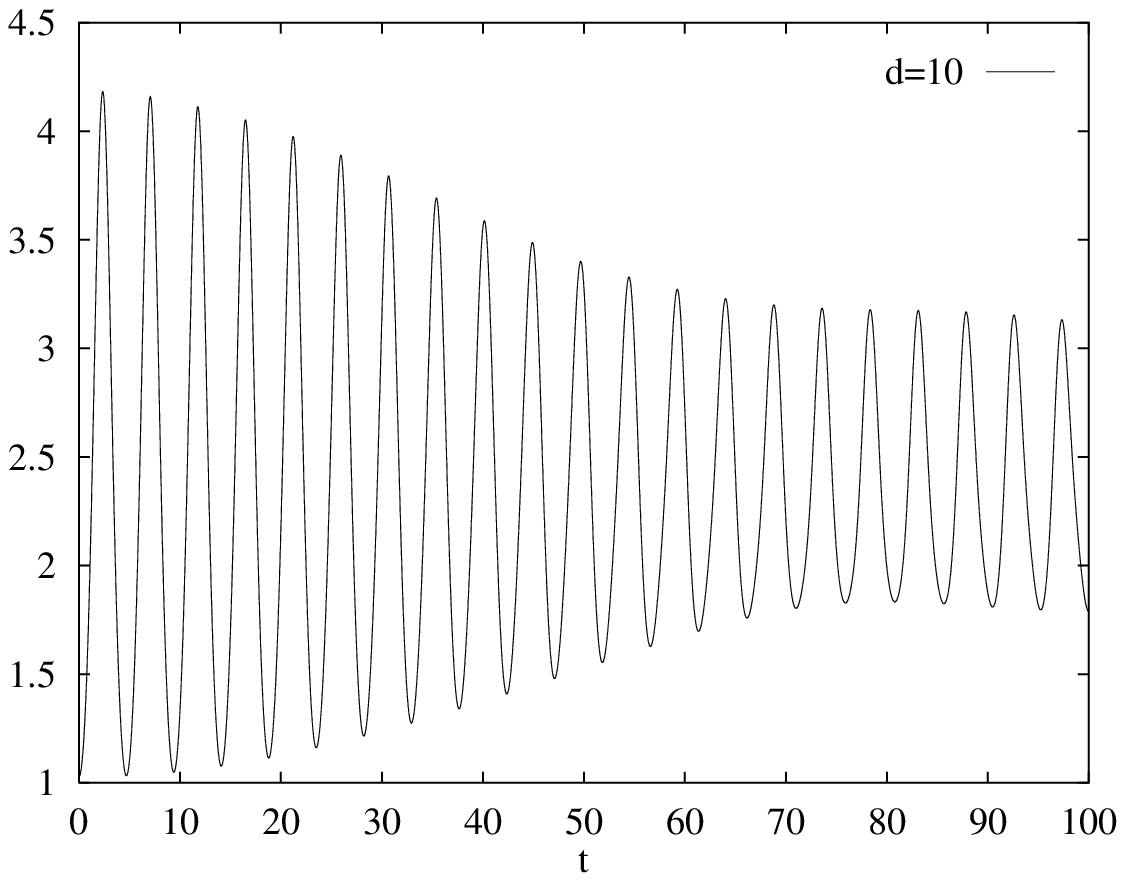}} 
   \epsfxsize = 2.2in 
   \subfigure[$d$=20]
             {\epsfbox{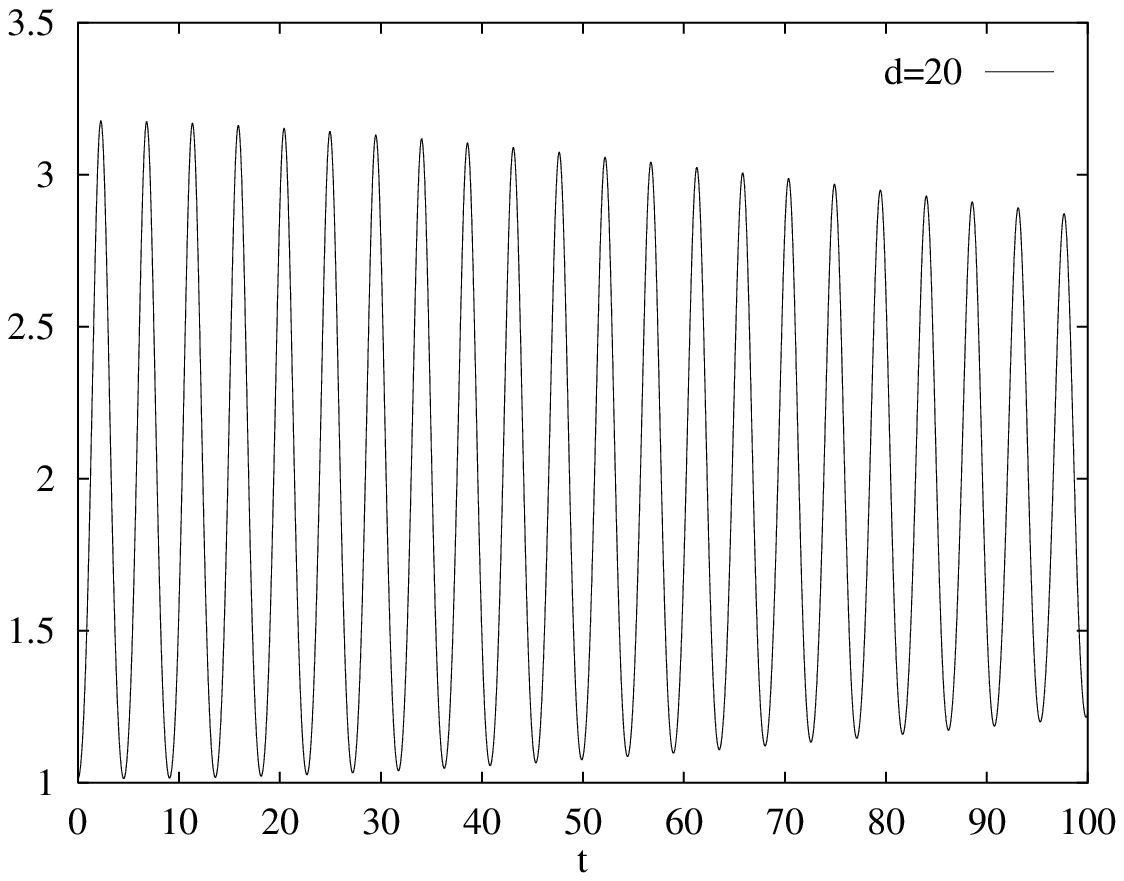}} 
   \caption{Long-time behavior of the exact solution for Gaussian
   initial conditions with $g_\infty=1$, $\bar{m}^2
   =2$, $G_{\infty}=1$.}  
   \label{fig:set1_exact}
\end{figure}

\begin{figure}
   \centering
   \epsfxsize = 2.2in
   \subfigure[$d$=1]
             {\epsfbox{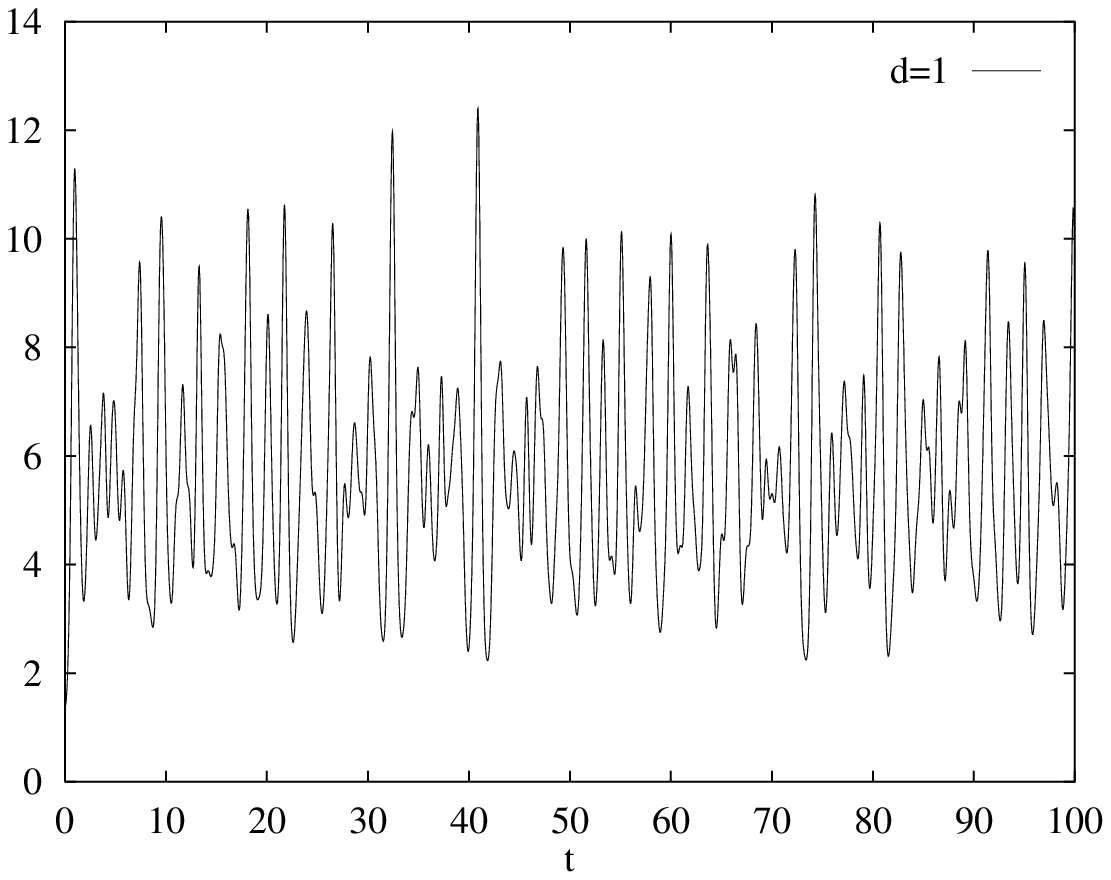}}
   \epsfxsize = 2.2in
   \subfigure[$d$=5]
             {\epsfbox{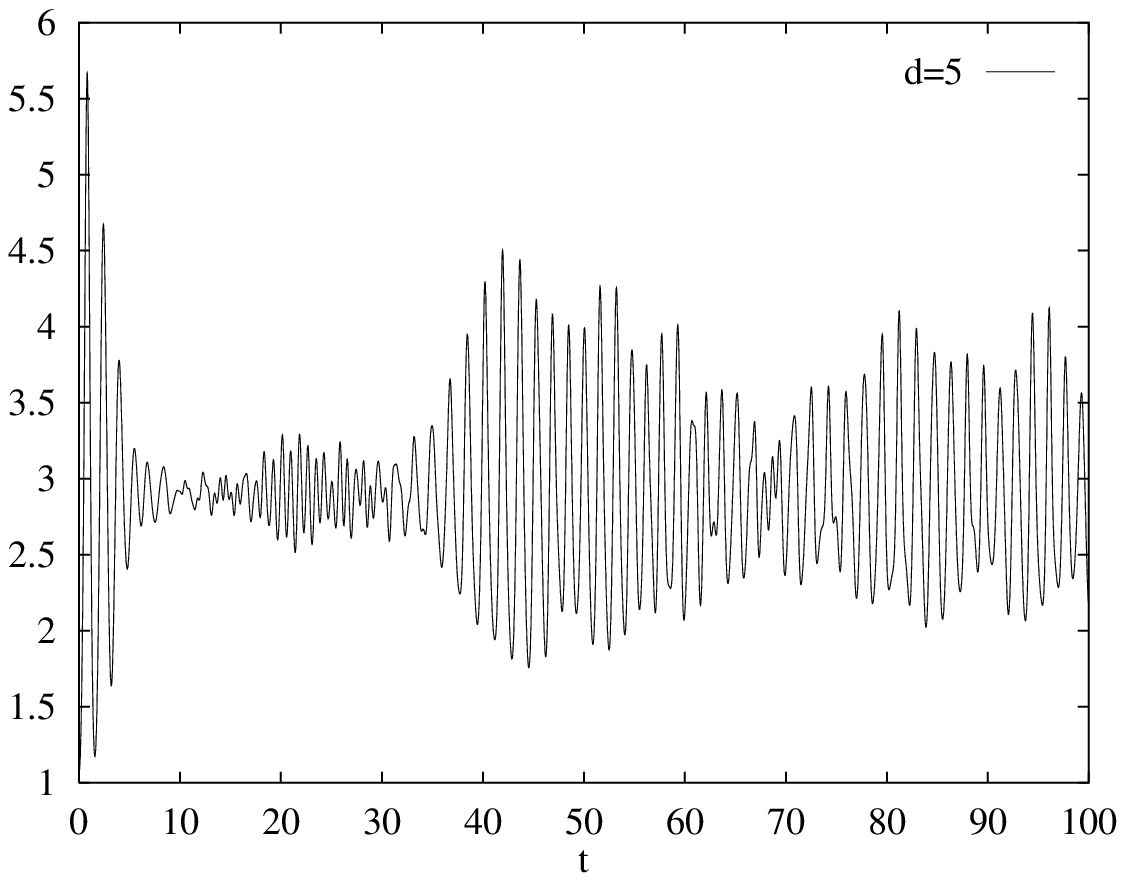}}
   \epsfxsize = 2.2in
   \subfigure[$d$=5]
             {\epsfbox{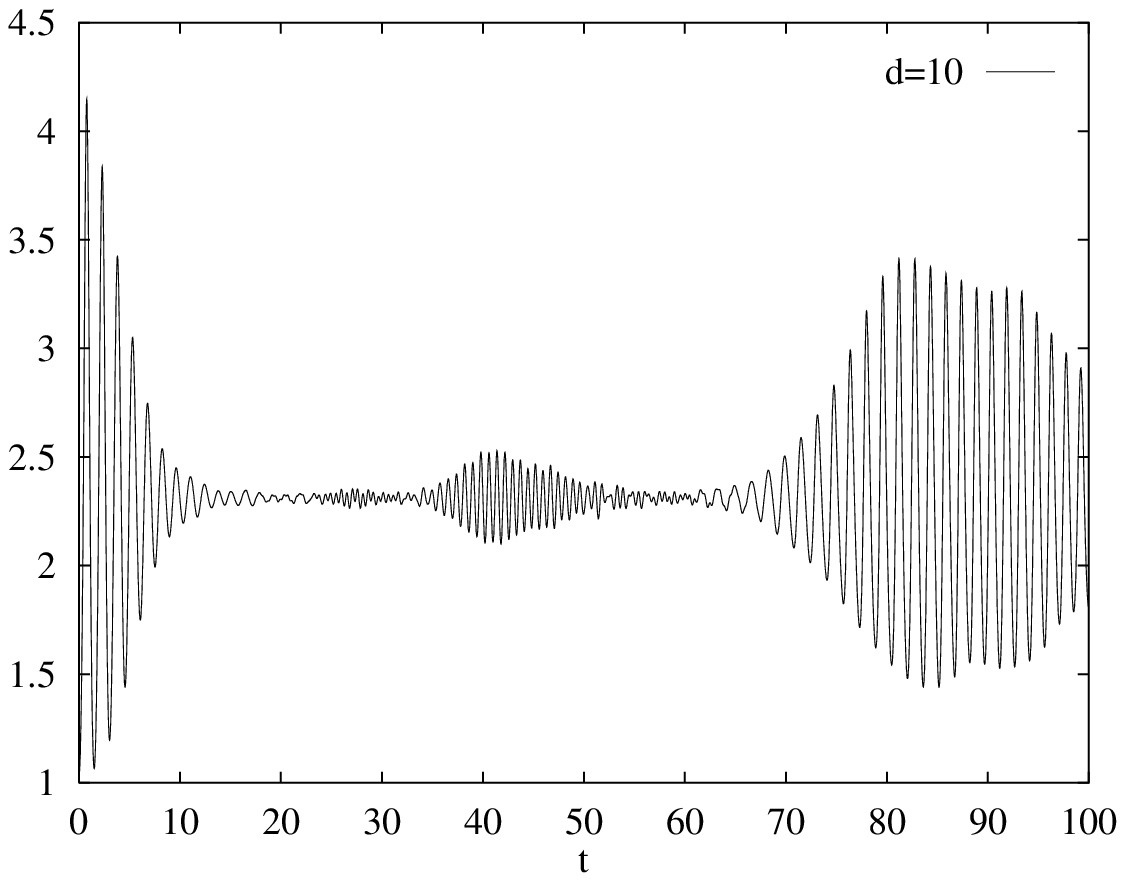}}
   \epsfxsize = 2.2in
   \subfigure[$d$=10]
             {\epsfbox{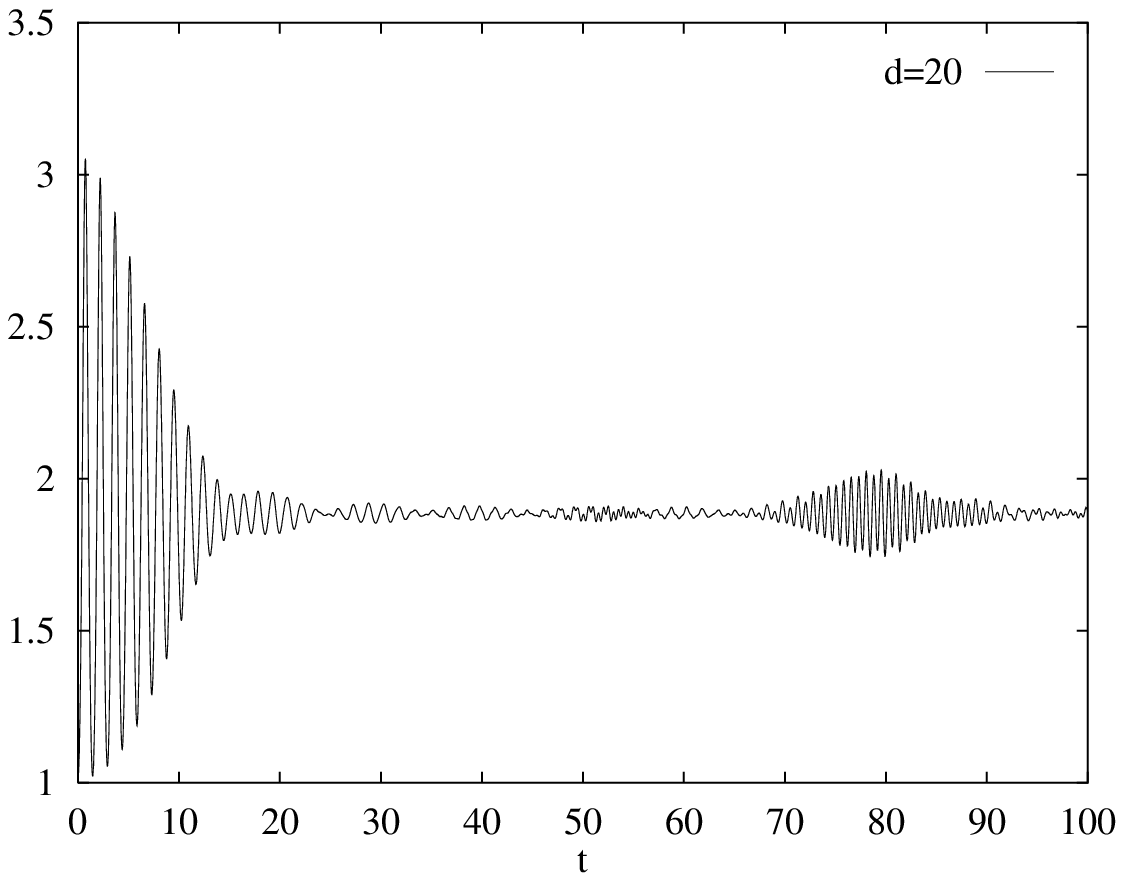}}
   \caption{Long-time behavior of the exact solution for 
            Gaussian initial conditions with $g_\infty=20$,
            $\bar{m}^2 =21$,$G_{\infty}=1$.}
   \label{fig:set2_exact}
\end{figure}

The few examples that we consider in this paper show a surprising
variety of long-time behaviors.  In
Figs.~\ref{fig:set1_exact}-\ref{fig:set2_exact} we show the evolution
of $G_2= \langle r^2/d \rangle$ over a time interval sufficiently long
to capture the characteristic behavior of that particular solution.

In Fig.~\ref{fig:set1_exact} we notice that as we increase $d$ we
gradually approach the limiting form of oscillation in an harmonic
well.  In the case $d=1$, $\bar m^2=2$, (Fig.~\ref{fig:set1_exact}a
$G_2$ is dominated by the contribution of the lowest frequency and so
appears as a perturbation of a simple oscillation with frequency equal
to the energy difference between the ground and first excited states.
The solution itself contains one dominant peak with a few smaller
peaks. As we increase $d$ the potential gets more harmonic and simple
harmonic motion of $G_2$ is observed.  At larger $g_{\infty}$, as seen
in Fig.~\ref{fig:set2_exact}, more complicated behavior can be seen.

In the case $d=1$, $\bar m^2=21$, (Fig.~\ref{fig:set2_exact}b) there
is a significant modulation of the amplitude, so that at regular
intervals there is a much reduced oscillation of $G_2$.  The solution
typically has several (2-3) peaks of similar size.

In the other cases in Fig.~\ref{fig:set2_exact}, there is always
significant amplitude modulation so that there may be brief or in some
cases ($d=10$ and $20$, $\bar m^2=21$), extended periods when the
amplitude of the oscillation is an order of magnitude less than the
initial oscillation.  The oscillation always returns to a significant
fraction (50-90\%) of the initial oscillation, even after extended
periods of reduced amplitude.  The pattern of amplitude modulation can
be quite complex.  The solutions in general have a number of major
peaks, from a minimum of 3 up to 12, indicating the contributions from
higher eigenstates.  However, in some cases the coherence of the
initial Gaussian state may be maintained for a significant length of
time.  For example, in the case $d=20$, $\bar m^2=2$, the coherent
state is maintained approximately for 10 oscillations at the
fundamental frequency. For this large value of $g$ we expect that one
needs to get to very large values of $d \ge 100$ before the asymptotic
behavior seen in Fig.~\ref{fig:set1_exact}d will be reached.

\section{Conclusions}

In this paper we have studied the behavior of $G_2 = \langle r^2/d
\rangle$ for Gaussian and non-Gaussian initial data for the quantum
roll in $d$ dimensions.  We found that as $d \rightarrow \infty$,
$G_2$ became more harmonic in behavior as expected.  We compared the
exact numerical data with two variational approximations (Gaussian and
Post Gaussian) for the wave function as well as the leading and next
to leading order approximations in the large $d$ expansion for the
Green's functions.  We found that all these approximations converged
to the exact result as $d \rightarrow \infty$.  At short times (less
than 2 oscillations) the next to leading order large $d$ approximation
was the most accurate one.  However when this approximation started
diverging from the exact result, it broke down in a serious fashion in
that $G_2$ became negative. This unexpected result is related to the
fact the large $d$ expansion for the expectation values may not
necessarily correspond to a positive definite probability when
truncated at any finite order in $1/d$.  At lowest order, however the
approximation is equivalent to a Gaussian density matrix and does not
have this defect.  Variational approximations to the wave function by
their very nature never have this defect. Recently \cite{ref:bett} we
have seen that in a related approximation, namely truncating the equal
times connected Green's function at the four point function level,
$G_2$ is not positive definite. Thus we feel that truncating the
heirarchy of Green's functions in various ways might always have this
problem, apart from the Gaussian case.  This leads us to suspect that
a higher order variational approach will be necessary if we want to
include scattering corrections to mean field theory and also look at
long time behavior. This will be the subject of a future paper. For
astrophysical applications of the quantum roll problem, one is only
interested in getting the correct result to the bottom of the
potential, after which particle production leads to dissipative
effects as one oscillates at the correct new minimum.  For this
purpose, keeping the next to leading order in large $N$ definitely
improves the mean field result.

\section{Acknowledgements}

The work by B.M. and J.F.D. at UNH is supported in part by the
U.S. Department of Energy under grant DE-FG02-88ER40410.  B.M. and
J.F.D thank the Theory Group (T-8) at LANL for hospitality during the
course of this work.  J.F.D. would also like to thank the Nuclear
Theory Center at Indiana University for hospitality during some of
this work.

%
%
\appendix

\section{Numerical Approach}

\subsection{Analytical Preliminaries}

The time-dependent Schr\"odinger equation
\begin{equation}
   \left  \{ 
      - \frac{1}{2} \frac{\partial^2}{\partial r^2} 
      + U(r) 
   \right \} \phi = i \frac{\partial \phi}{\partial t}
\end{equation}
where $U(r)$ is given by (\ref{eq:Uofr}), 
\begin{equation}
   U(r) = \frac{(d-1)(d-3)}{8r^2} 
      + \frac{g}{8d}(r^2-r_0^2)^2 \>, 
\end{equation}
is solved by the method of eigenfunction expansion.  Taking a Fourier
transform in time, we have the eigenvalue problem
\begin{equation}
   -\frac{1}{2}\Phi'' + U \Phi = E\Phi
\end{equation}
on the half-line $r>0$.  The boundary conditions are finiteness at the
origin and the normalization condition,
\begin{equation}
   \int_0^{\infty}|\Phi(r)|^2 \, {\rm d}r = 1 \>.
\end{equation}
The numerical solution of this eigenvalue problem is described in the
next section.  The value of the wave function at $t = 0$ is given by:
\begin{equation}
   \phi(r,0) = \phi_0(r) = A r^{(d-1)/2} \exp( -r^2/4G ) \>,
\end{equation}
where $A$ is a normalization constant depending on $G$ and $d$.  
So the initial-value problem is solved by the expansion
\begin{equation}
   \phi(r,t) = \sum_{j=0}^{\infty} c_j \exp(-iE_jt)\Phi_j(r)
\end{equation}
where the coefficients are determined by quadrature
\begin{equation}
   c_j = \int_0^{\infty}\phi_0(r) \, \Phi_j(r) \, {\rm d}r \>.
   \label{eq:Avii}
\end{equation}
To evaluate a particular moment of the solution without determining the
full spatial solution, we take the moment of the expansion.  Thus we may
calculate the second moment as follows
\begin{equation}
   \langle r^2 \rangle = 
      \sum_{j,k=0}^{\infty} a_{j,k} \, c_j c_k \, 
      \cos(E_j-E_k)t  \>, 
\end{equation}
where
\begin{equation}
   a_{j,k}= \int_0^{\infty} 
      r^2 \, \Phi_j(r) \, \Phi_k(r) \, {\rm d}r
   \>.
   \label{eq:Aix}
\end{equation}

\subsection{Numerical Solution of Eigenvalue Problem}

\subsubsection{Discretization}

The ordinary differential equation (2) is discretized using a compact
finite difference scheme of second-order accuracy.  This scheme
allowed the easy use of a nonuniform mesh near the origin, which was
needed especially in the lower-dimensional cases ($d<10$) to acheive
the desired accuracy. The nonuniform grid $\{r_n\}$ was created by
transforming the independent variable by the formula
\begin{equation}
   r = \frac{s^b } { (1 + s)^{(b+1) } }
\end{equation}
where $b = 2/(d-3)$ and then using a uniform grid [$h$, $2h$,
$\ldots$, $L$] in the variable $s$.  The discretized equations are
then
\[
   \frac{u_{n+1}-u_{n}}{r_{n+1}-r_{n}} = 
   \frac{w_{n+1}+w_n}{2},
\]
and
\begin{eqnarray*}
   && \qquad\qquad\qquad
   -\frac{1}{2}\left(\frac{w_{n+1}-w_{n}}{r_{n+1}-r_{n}}\right) 
  + \\ &&
     \left  ( 
        \frac{ U(r_{n+1})+U(r_n) }{ 2 } - E 
     \right )
     \left  ( 
        \frac{ u(r_{n+1})+u(r_n) }{ 2 }
     \right )
   = 0 \>.
\end{eqnarray*}
The boundary conditions at the singular points $0$ and $\infty$ were
imposed by specifying $w=0$ at the leftmost grid point and $u=0$ at the
rightmost grid point.  The parameters $h$ and $L$ were adjusted until
sufficient accuracy was acheived, as described in the next section.

\subsubsection{Eigensolver}

In matrix notation we have $A \, x = E \, B \, x$, where $x$ is a
vector containing the $u$ and $w$ values at the grid points.  This is
a generalized eigenvalue problem and the matrix $B$ is singular.  We
can formally convert this to a regular eigenvalue by inverting $A$,
(although we will never actually invert $A$ in practice.)  We then
have $A^{-1} B \, x = \lambda \, x$, where $\lambda = E^{-1}$.  This
eigenvalue problem was solved for a specified number (16 usually) of
the largest eigenvalues by the method of Arnoldi
factorization~\cite{ref:GvL}, which is an iterative method and so is
very fast provided matrix vector multiplication can be carried out
quickly.  MATLAB routines where used for the Arnoldi factorization as
well as an incomplete L-U factorization of the matrix $A$ that was
used to quickly evaluate the matrix-vector products $A^{-1}Bx$.

Convergence of the solution was confirmed by testing the orthogonality of
the eigenvectors and by computing the energy of the eigenfunctions from
the formula
\[
   E = 
   \int_0^{\infty} 
   \left  \{ 
      \frac{1}{2} \left| 
                     \frac{\partial \Phi(r)}{\partial r}
                  \right|^2 + 
      V(r)\, |\Phi(r)|^2 
   \right \} \, {\rm d}r  \>.
\]
The size of the interval, as determined by $L$, and the mesh spacing,
as determined by $h$, that were required to acheive an accuracy of
$10^{-3}$ was highly dependent on the number of eigenvalues being
determined and to a lesser extent on the dimension $d$.  At most 4096
grid points were required in the worst case of 32 eigenfunctions with
$d=20$ --- usually 1024 grid points were sufficient.

\subsection{Solving the Initial-Value Problem}

To solve the initial-value problem, the only remaining step is to
evaluate the integrals (\ref{eq:Avii}).  This is done by using the
trapezoidal rule.  The accuracy of the quadrature was verified by
reconstructing the initial condition and the mesh refined until an
accuracy of $10^{-3}$ was acheived.  The integrals (\ref{eq:Aix}) were
evaluated similarly.

%
%
%
%

%
%
%
%
%
%
%
%
%
%
%
%
%
%
%
%
%
%
%
%
%
%
%
%
%
%
%
%
%
%
%
\newpage 
\begin{table}
\begin{tabular}{c|ccccc}
   $d$   &       $g$ &     $G$ &   $m^2$ & $\tilde r$ & $r_0$   \\ 
   \hline
    1    &    0.2320 &  1.3760 &  2.0000 &  0.9359    & 2.9359  \\
    5    &    0.5628 &  1.0572 &  1.9805 &  2.1877    & 4.1739  \\
   10    &    0.7266 &  1.0269 &  1.9315 &  3.1255    & 5.0638  \\
   20    &    0.8728 &  1.0129 &  1.8596 &  4.4451    & 6.2765  \\
$\infty$ &    1.0000 &  1.0000 &  1.2500 & $\sqrt{d-1}$ &       \\
\end{tabular}
\caption{Initial values for $\bar{m}^2 = 2.0$, $g_{\infty} = 1.0$}
\label{table:table1}
\end{table}

\begin{table}
\begin{tabular}{c|ccccc}
   $d$   &       $g$ &     $G$ &   $m^2$ & $\tilde r$ & $r_0$   \\ 
   \hline
    1    &    2.4363 &  1.3760 & 21.0000 &  0.9359    & 2.9359  \\
    5    &    5.9798 &  1.0572 & 20.9805 &  2.1877    & 4.1865  \\
   10    &    7.9590 &  1.0269 & 20.9315 &  3.1255    & 5.1199  \\
   20    &   10.0209 &  1.0129 & 20.8596 &  4.4450    & 6.4305  \\
$\infty$ &   20.0000 &  1.0000 & 20.2500 & $\sqrt{d-1}$ &       \\
\end{tabular}
\caption{Initial values for $\bar{m}^2 = 21.0$, $g_{\infty} = 20.0$}
\label{table:table2}
\end{table}
%
%
\end{document}